\documentclass[twocolumn]{revtex4-1}
\usepackage[utf8]{inputenc}

\usepackage{amsmath}
\usepackage{amssymb}
\usepackage{graphicx}
\usepackage{caption}
\usepackage{epstopdf}
\usepackage[section]{placeins}
\usepackage{fancyhdr}
\usepackage[left=2.0cm,right=2.0cm,top=2.3cm,bottom=2.2cm]{geometry}
\usepackage{xcolor}
\usepackage{mhchem}

\date{\today}

\begin{document}

\author{Rina Ibragimova$^1$}
\author{Zhong-Peng Lv$^1$}
\author{Hannu-Pekka Komsa$^{1,2}$}
\affiliation{
$^1$Department of Applied Physics, Aalto University, 
Finland
}
\affiliation{
$^2$Microelectronics Research Unit, University of Oulu, 
Finland}

\title{First principles study of stability of MXenes under electron beam}

\begin{abstract}
Interactions of two-dimensional MXene sheets and electron beam of
(scanning) transmission electron microscope are studied via
first-principles calculations.
We simulated the knock-on displacement threshold for Ti$_3$C$_2$ MXene sheet via \emph{ab initio} molecular dynamics simulations and for five other MXenes (Ti$_2$C, Ti$_2$N, Nb$_2$C, Mo$_2$TiC$_2$, and Ti$_3$CN)
approximately from defect formation energies.
We evaluated sputtering cross section and sputtering rates,
and based on those the evolution of the surface composition.
We find that at the exit surface and for ``low'' TEM energies H and F sputter at equal rates, but at ``high'' TEM energies the F is sputtered most strongly. In the enter surface, H sputtering dominates.
The results were found to be largely similar for all studied MXenes, and although the displacement thresholds varied between the different metal atoms the thresholds were always too high to lead to significant sputtering of the metal atoms.
We simulated electron microscope images at the successive stages of sputtering, and found that while it is likely difficult to identify surface groups based on the spot intensities, the local contraction of lattice around O groups should be observable.
We also studied MXenes encapsulated with graphene and found them to provide efficient protection from the knock-on damage for all surface group atoms except H.
\end{abstract}

\maketitle

\section{Introduction}

MXenes are a class of two-dimensional materials of transition
metal carbides and nitrides, with the chemical formula M$_{n+1}$X$_n$T$_x$
\cite{Naguib11_AM,Ghidiu14_Nat,Anasori17_NRevMats}.
These materials are obtained \emph{via} selective etching of the layered bulk precursor phases using, e.g., hydrofluoric acid (HF) \cite{Anasori17_NRevMats, rev_synth19}, which results in the surface sites passivated by functional groups T$_x$ from the solution, where T$_x$ predominantly consists of O, OH, and F.
MXenes possess many beneficial properties, such as good electrical conductivity, hydrophilicity, flexibility, mechanical strength, consisting of abundant elements, stability in solution and the ease of synthesis in large batches.
In particular, the combination of these has made them suitable for many applications such as batteries and supercapacitors, electromagnetic interference shielding, sensors, wearable devices
\cite{Ghidiu14_Nat,Zhang17_AM,Cao18_ACSNano,Hantanasirisakul16_AEM, Shahzad1137}. 
Importantly, the surface functional groups as well as defects in the MX backbone can have a significant effect, either beneficial or detrimental, on the material properties.
Thus plenty of effort has been devoted to study them via XPS, NMR, X-ray and neutron scattering, Raman spectroscopy, and
(scanning) transmission electron microscopy [(S)TEM].
(S)TEM appears particularly suited for the study of defects as it can provide direct structural information, and has been successfully employed with other 2D materials \cite{MeyerNL09-BN,JinPRL09-BN,Krivanek10nat,Komsa12_PRL}.

(S)TEM images of the defects on Ti3C2 MXenes were reported in
\cite{Karlsson15_NL,Sang16_ACSNano,Persson18_2DM,Sang2018},
and mostly show Ti-related defects: vacancies and adatoms.
Unfortunately, C, O, F, and H atoms are difficult to identify reliably, since STEM signal is proportional to the atomic number as $\sim Z^{1.7}$, 
and thus Ti atoms dominate the signal. 
While the O, OH, and F atoms cannot be seen directly, the brighter areas
were assigned to regions with higher concentration of O atoms, which seemed
to agree with EELS \cite{Karlsson15_NL,Persson18_2DM}.
In addition, 
Sang et al. \cite{Sang16_ACSNano} observed correlation between the etching conditions and vacancy concentration, and also clustering of vacancies.




The interaction of the relativistic electron and the sample can lead to damage via several different mechanisms \cite{Susi19_NatRevPhys}.
Elastic collision between electron and nucleus is called knock-on mechanism.
In inelastic collision, energy is lost to electronic excitations leading to direct bond breaking (radiolysis), heating, or charging.
In addition, the beam can crack gas molecules or contamination, and these radicals can lead to chemical etching \cite{Leuthner19_UM}.
In the case of graphene, knock-on dominates and heating and charging effects can be ignored due to very high electrical and thermal conductivity. 
In the case of BN, the charging effects can be significant. 
Radiolysis is known to be important organic compounds, but plays a minor role in conducting samples due to short excitation lifetime.
Semiconductors, such as TMDs fall somewhere in between \cite{Komsa12_PRL,AlgaraSiller13_APL}.
Also, in the case of MXenes, due to the high electrical conductivity, we expect radiolysis and charging effects to play minor role. The thermal conductivity is also reasonably high.
In knock-on mechanism, the heavier atoms are more stable under the beam.
All atoms in Ti$_3$C$_2$T$_x$ are relatively light and thus susceptible to sputtering.
In Ref.\ \cite{Sang2018}, Sang et al studied Ti dynamics and observed Ti displacement and hole growth, but also formation of thicker Ti$_{n+1}$C$_n$ layers.
According to Zhang et al. \cite{Zhang19_CM}, prolonged irradiation parallel to the layers led to removal of H and ``repartitioning'' of Ti and O atoms between the MXene layers.
Although the lighter atoms are expected to be sputtered, since they are not directly seen, the microscopic details have remained elusive.
To this end, atomistic simulations could prove highly useful in providing the missing details.

Here, we present a first principles study on the stability of MXenes under electron beam.
We carried out \emph{ab initio} molecular dynamics (AIMD) simulations to determine displacement thresholds and sputtering cross sections of functional groups and Ti atoms of Ti$_3$C$_2$T$_x$ monolayers.
We focus on the knock-on, since we expect it to dominate for the above-mentioned reasons, but also because it is most straightforward simulate reliably using first principles simulations.
Based on these we can estimate the order (and rate) at which each functional group is removed from the top and bottom surfaces.
We also simulate (S)TEM images from the structures at successive stages of sputtering.
For several other MXenes, we evaluate the displacement thresholds using unrelaxed vacancy formation energies, since AIMD is computationally demanding and this approximation is found to work well.
We also consider protecting the MXenes sheets by sandwiching it with graphene that has been found to be resistant to damage from electron irradiation.
Finally, the results are compared to the existing literature.

\section{Methods}

Density-functional theory calculations were used to model the electron-beam interaction with single MXene sheets.
All calculations were performed using the projector augmented wave formalism as implemented in the simulation package VASP \cite{kres1,kres2}.
We adopt Perdew-Burke-Ernzerhof for solids (PBEsol) exchange correlation functional \cite{PBEsol} for the all calculations. For the heterostructure with graphene, we also include DFT-D3 van der Waals corrections \cite{DFT-D3}.
The optimal plane-wave cutoff energy was chosen as 500 eV according to the convergence test. The k-points set of 3$\times$3$\times$1 was chosen as optimal for monolayer calculations with the 4x4x1 size of supercell. 

As we found previously in the case of Ti$_3$C$_2$, composition of surface functional groups at given pH and work function can be a mixture of F, O, and OH 
\cite{Ibragimova19_ACSNano}. 
Here, we employ a 4$\times$4 supercell special quasi-ordered structures with a composition O$_{0.5}$F$_{0.25}$OH$_{0.25}$, as constructed in Ref.\ \cite{Ibragimova19_ACSNano}.
The heterostructure of Ti$_{3}$C$_{2}$-graphene was constructred by placing the 4$\times$4 MXene supercell on 5$\times$5 supercell of graphene and the lattice constant fixed to that of MXene sheet.
Defects in the other MXene sheets are also modeled using a 4$\times$4 supercell.

\begin{figure}[ht]
\centering
\includegraphics[width=\columnwidth]{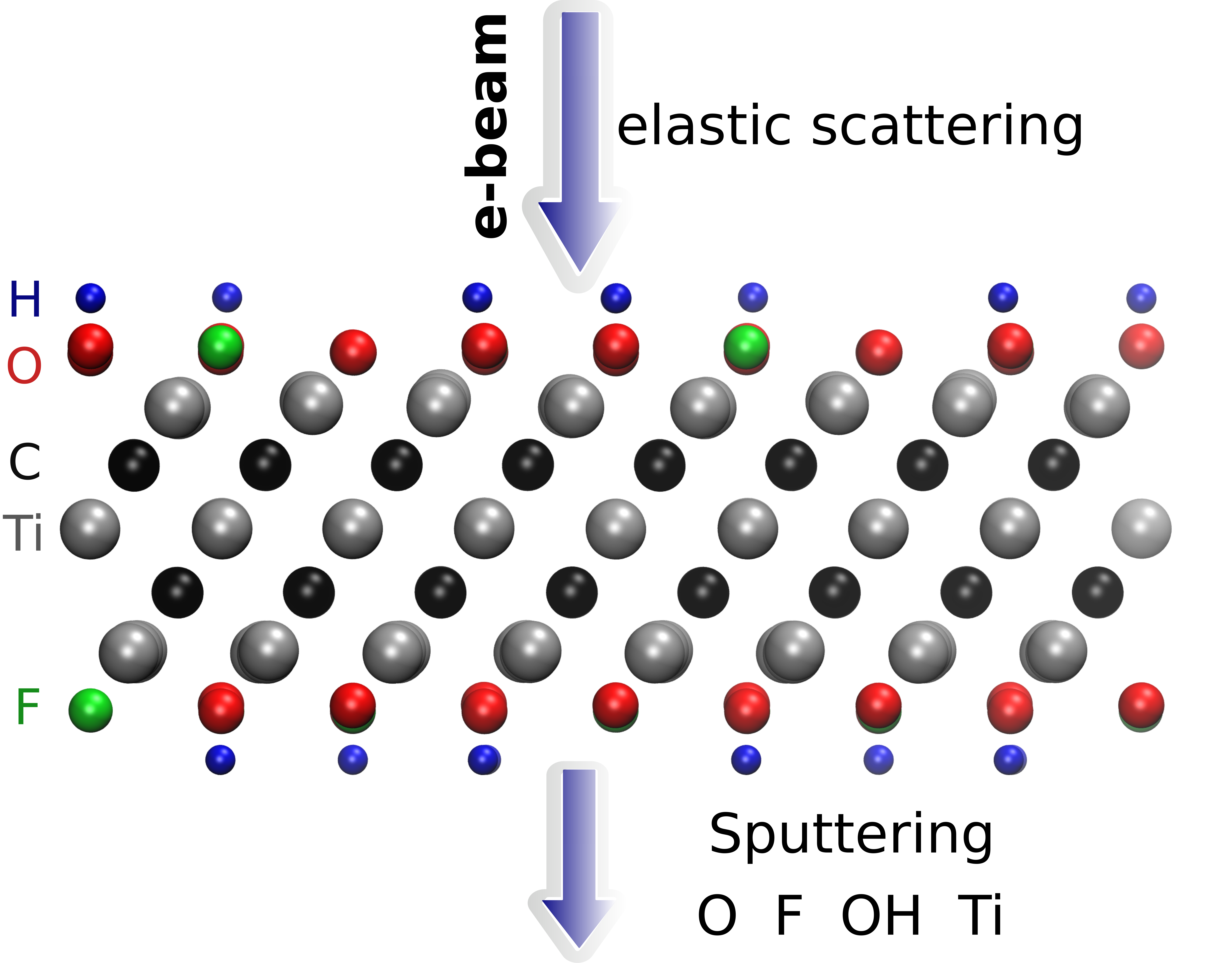}
\caption{\label{fig:setup}
Illustration of the simulation setup used for \emph{ab initio} MD calculations of Ti$_3$C$_{2}$O$_{0.5}$F$_{0.25}$OH$_{0.25}$. Adopted convention for the direction of the electron beam is also shown.
}
\end{figure}

The threshold energy for sputtering atoms was determined by running a series of \emph{ab initio} molecular dynamics (AIMD) simulations. 
Assuming that momentum transfer from the electron to the atom is instantaneous and the collision is fully elastic, we can then use the energy and momentum  conservation principle. The initial kinetic energy transferred to the atom is increased until 
we find a minimal energy needed to sputter the atom from lattice.
The calculations are performed with a step of kinetic energy 0.1 eV for those processes with low displacement threshold and with a step of 1 eV for those with high displacement threshold (above 20 eV). 
The computational setup is illustrated in Fig. \ref{fig:setup}. We adopt a convention where the electrons enter from the top and exit from the bottom of the sheet.
We used a small time step of 0.5 fs for most of the molecular dynamics calculations and 0.1 fs timestep for modelling H sputtering.

Sputtering cross sections are calculated using McKinley-Feshbach formalism
\cite{McKinley48_PR}, which is generally valid (i.e., equal to Mott formula) at $Z<29$ and thus valid for all atoms in our Ti$_3$C$_2$T$_x$ systems.
The effect of finite ion velocities is accounted for as described in Ref.\ \cite{Meyer12_PRL} and adopting Maxwell-Boltzman velocity distribution.

The sputtering rate of atoms can be found by multiplication of cross section by an current density of a microscope, assuming that 1 electron sputters out 1 atom, $S = \sigma J$.
The reported electron microscopy studies of MXenes \cite{Sang16_ACSNano,Karlsson15_NL,Persson18_2DM,Persson18_2DM,Zhang19_CM}
have used electron energies 60, 80, 100 and 300 keV and beam currents
of 10--100 pA, where reported. 
Usually larger current is needed at lower electron energies. 
Assuming STEM configuration with beam current of 15 pA focused on an area of one unit cell (or one surface site, about 9 {\AA}$^2$), we obtain a proportionality factor of 0.1, i.e., for 100 barn cross section a sputtering rate of about 10 atoms/s.
We expect that our estimate for the rate is likely on the lower bound.
The rate equations for determining the evolution of composition with time are given in the Supplementary information \cite{supp}.


For the other MXenes, the sputtering threshold is evaluated only using the unrelaxed defect formation energy approach, which was found to work well in the case of ``rigid'' 2D materials and when the sputtered atom's trajectory is unobstructed \cite{Komsa12_PRL}. 

STEM image simulations were carried out using DrPROBE software \cite{DrProbe}
at acceleration energies of 60 and 300 keV and using probe size 0.04 nm (half-width-half-maximum) \cite{Sang16_ACSNano,Sasaki12_Mic}.
Other simulations parameters are convergence semi-angle 21.5 mrad, spherical aberration 200 nm, chromatic aberration 1.5 mm, 3-fold astigmatism $< 30$ nm, axial coma $< 30$ nm, 4-fold astigmatism 500 nm, star aberration 500 nm, HAADF inner detector angle 80 mrad, and HAADF outer detector angle 250 mrad.
%

\section{Results}

\subsection{Ti$_3$C$_2$ under electron beam}

The calculated displacement threshold energies, as well as the corresponding electron energies, are given in Table \ref{tab:en_mix}
and in Fig.\ \ref{fig:thresholdbar}.
We have considered displacement of O and F ions in the case of O and F functional group, both O and H separately in the case of OH group (denoted respectively as OH and H in Table \ref{tab:en_mix}), and Ti atoms in the bare (unterminated) surface.
Note, that we do not consider the angular momentum transfer and secondary collisions which will increase the number of sputtered atoms. 


\begin{table}[h] 
\centering
\caption {Calculated sputtering threshold energy (transferred from electron to atom) T$_{k}$ and the corresponding TEM acceleration voltage E$_{el}$ for Ti$_3$C$_2$O$_{0.5}$F$_{0.25}$OH$_{0.25}$ surface.} \label{tab:en_mix} 
\begin{ruledtabular}
\begin{tabular}{lcccc}
    &\multicolumn{2}{c}{Top}& \multicolumn{2}{c}{Bottom} \\
&T$_{k}$ (eV)&E$_{el}$ (keV)&T$_{k}$ (eV)&E$_{el}$ (keV)\\
\hline
O&32 &200& 9.5&65\\
F&22&165&5.5 &46\\
H&4.7&2.2 &3.8&1.7\\
OH&32&200 & 10.6& 72 \\
Ti&25&391&15.9&270\\
\end{tabular}
\end{ruledtabular}
\end{table}

\begin{figure*}[ht!]
\includegraphics[width=\textwidth]{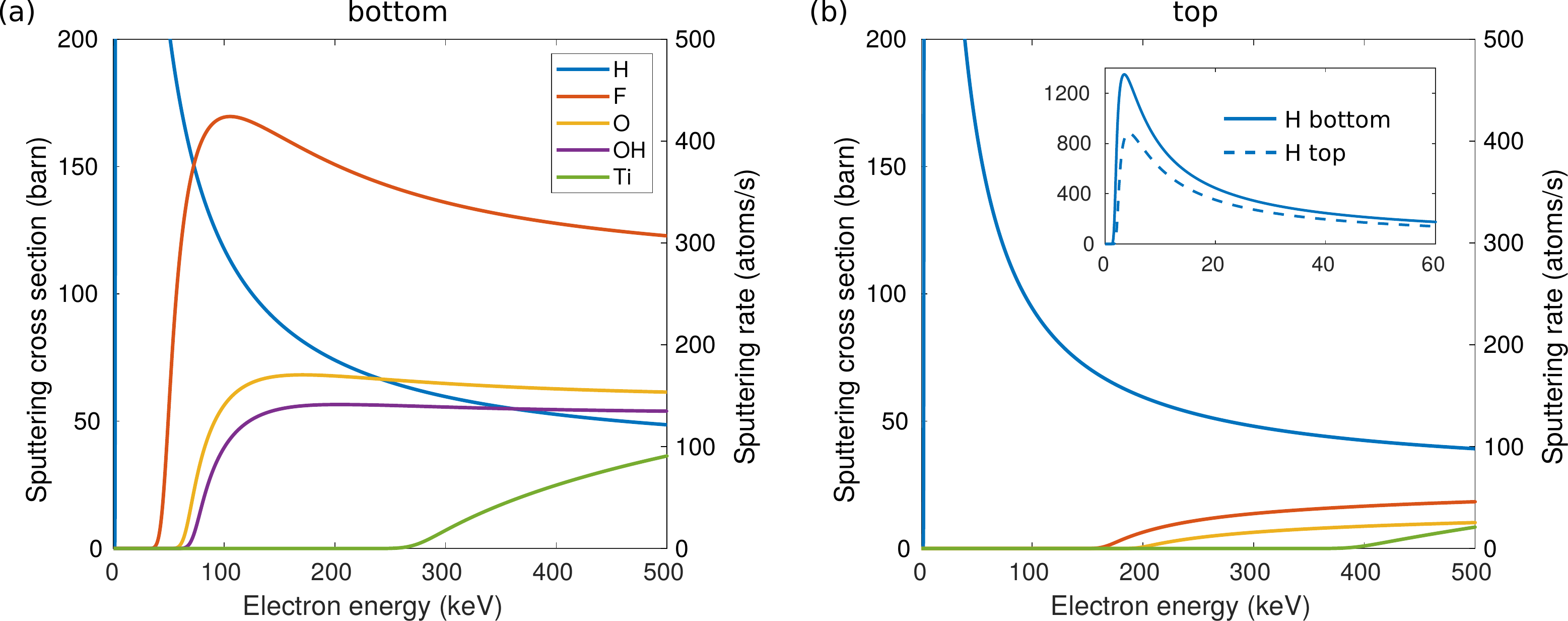}
 \caption{\label{fig:crosssection}
 Sputtering cross section for H, F, O, OH and Ti and corresponding sputtering rates depending on TEM electron energy from the bottom (a) and top (b)
 surfaces of Ti$_3$C$_2$O$_{0.5}$F$_{0.25}$OH$_{0.25}$.
 The inset zooms in to the H cross sections at low electron energies.}
\end{figure*}

First focusing on the bottom side,
threshold energies for sputtering of functional groups, and consequently the electron energies, are rather small: H sputters out from the surface at above 1.7 keV, F is sputtered at 46 keV, O at 68 keV, and finally OH at 72 keV.
The threshold energies were found not to vary significantly with composition. Slightly lower energies were obtained for sputtering atoms from pure terminated surface (e.g. about 9.2 eV for pure O-terminated surface)
and higher in case of O$_{0.5}$F$_{0.25}$OH$_{0.25}$ termination. 
We stress that the reported values correspond to sputtering of atoms. 
Various other processes may take place below the sputtering threshold
depending on the neighborhood of the sputtered atom(s).
In the case of 
O$_{0.5}$F$_{0.25}$OH$_{0.25}$ or pure surface termination, 
the nearly sputtered atoms fall back to the same position of the surface, mainly 
since we provide only momentum perpendicular to the surface. 
In case of O$_{0.5}$OH$_{0.5}$, 
the O atom falling back to surface can capture H atom from the neighboring OH sites, i.e., leading to diffusion of H,
or, at just below the sputtering threshold, even bond with two H atoms and form water molecule, which can desorb from or migrate on the surface.
%

On the other hand, Ti atoms (on a bare surface) are stable under the beam, up to 270 keV, and even then the sputtering rate is relatively low. Next to Ti vacancy, the threshold drops to 12 eV, but still relatively high.
This is consistent with the experimental observations of the Ti$_3$C$_2$ sheets
\cite{Karlsson15_NL,Sang16_ACSNano,Sang2018,Persson18_2DM}

On the top surface, displacement thresholds are much higher for all cases except H. 
The dominant sputtering process in our simulations is a straightforward ``bounce'' from the Ti$_2$C$_2$ backbone.
This is caused by the fact that we only carried out simulations with on-axis collisions, i.e., initial momentum perpendicular to the surface. We speculate that in the case of off-axis collision, the probability for e.g. formation of water or migration of functional groups is higher on the top surface than in the bottom surface.
Formation of water is an important aspect of MXenes, since many experimental studies report water presence between the layers even after drying \cite{Mashtalir13_MCP,Muckley17_ACSNano,Osti17_PRM}.

For a more quantitative estimate of the sputtering probabilities and the resulting surface composition, we first calculated the sputtering cross sections, as shown in Fig.\ \ref{fig:crosssection}.
Although sputtering H requires little kinetic energy and thus dominates at low electron energies, cross section at larger electron energies becomes smaller due to the low atomic number. 
At electron energies above 80 keV, F will be sputtered faster than H
and with probability about twice that for O.
The sputtering rates are shown on the right axis in
Fig.\ \ref{fig:crosssection}(a).
Only at 60 kV, some surface groups would remain at the bottom surface, but at $>100$ kV all functional groups should be quickly ($<1$ s) sputtered out and bottom surface becomes bare. 

\begin{figure*}[ht!]
\includegraphics[width=0.9\textwidth]{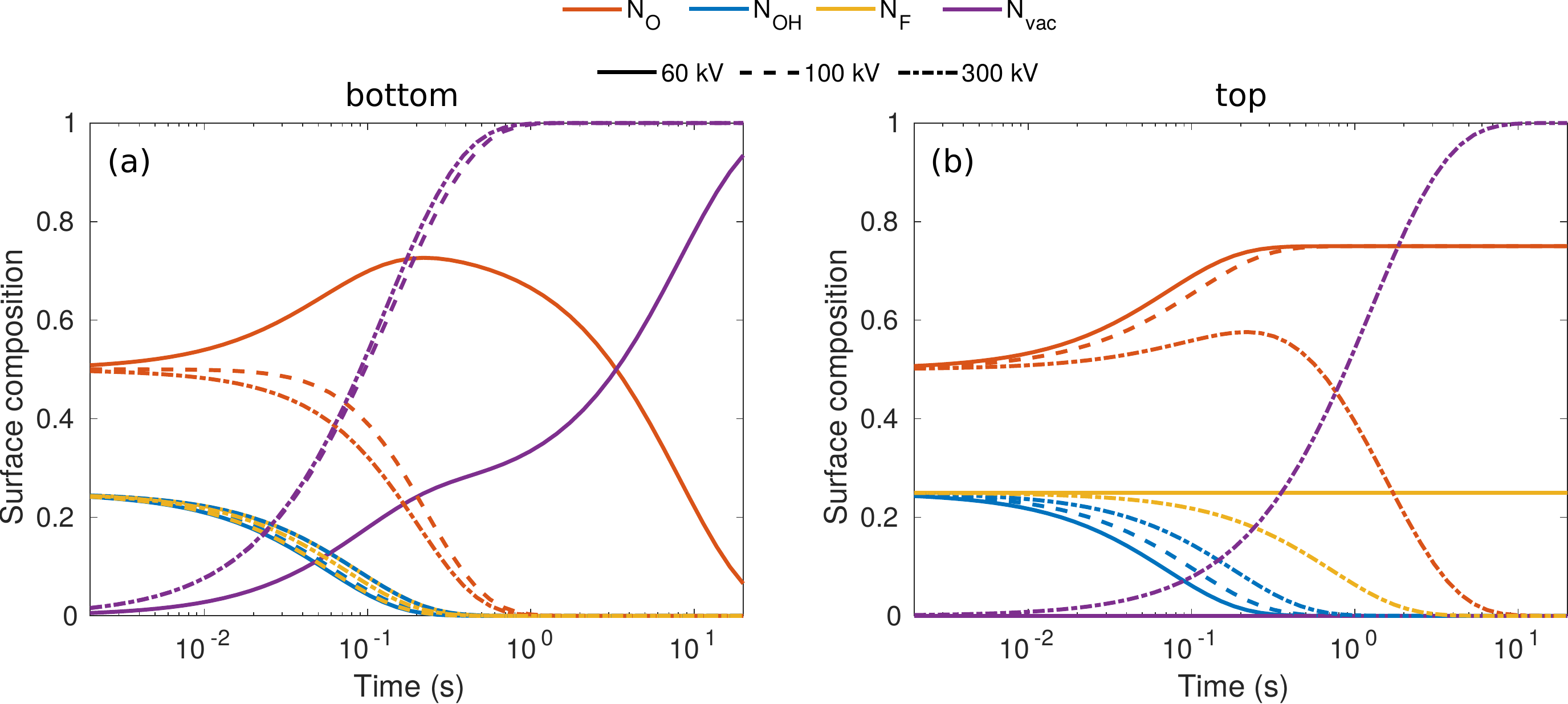}
\caption{\label{fig:rates}
Evolution of surface composition with time on the bottom (a) and top (b) surfaces for three typical acceleration voltages.
}
\end{figure*}

While the evolution of the surface composition at low voltages is clear, at 300 kV, sputtering rates for O, OH, and H are all similar and thus the composition evolution is less obvious. To this end, we solved the rate equations (see Supplementary information \cite{supp}).
The results for 60, 100, and 300 kV electron energies are shown 
in Fig.\ \ref{fig:rates}.
At 60 kV, the rapid H sputtering leads to conversion of all OH groups to O groups at both the top and bottom surfaces. In addition, F groups are quickly sputtered from the bottom surface, but not from the top.
At $t\approx 1$ s, the composition is close to O$_{0.75}$ at bottom surface and O$_{0.75}$F$_{0.25}$ at the top surface. This is followed by gradual O sputtering from the bottom surface, eventually leading to bare bottom surface, whereas the top surface remains covered by O and F.

At 100--300 kV, all surface groups are sputtered rapidly from the bottom surface, leading to bare surface already at $t=1$ s. 
On the top surface, 100 kV behavior is similar to 60 kV, but at 300 kV
both the F and O start sputtering out, although at a lower rate than from the bottom surface, leading to bare top surface at about $t=10$ s.

Overall, the sputtering rates are very high and thus we expect that bottom surface will quickly become clear of surface groups and also the top surface in certain conditions.
However, bare MXene surface is very reactive meaning that any residual gases in the chamber are likely to stick very effectively on it and thereby
``refunctionalizing'' it. Ambiguous functionalization naturally impedes  interpretation of the measured images.
On the other hand, if bare surface can be obtained, one could imagine exploiting it to introduce to the surface alternative groups, such as 
chalcogen and halogen group atoms \cite{Kamysbayev20_Sci,Qin20_JPCM} or even CO$_2$ \cite{Persson19_AM}.

\subsection{Surface group identification in TEM}

According to our results in the previous section, 
while the bottom surface is expected to become clear of functional
groups rather rapidly, 
at small doses the top surface should remain functionalized.
We simulated the TEM (and/or STEM) images for
the successive stages of the sputtering of functional groups
for the scenarios of low electron energy of 60 kV 
and high electron energy of 300 kV. 

\begin{figure*}[ht!]
\centering
\includegraphics[width=0.9\textwidth]{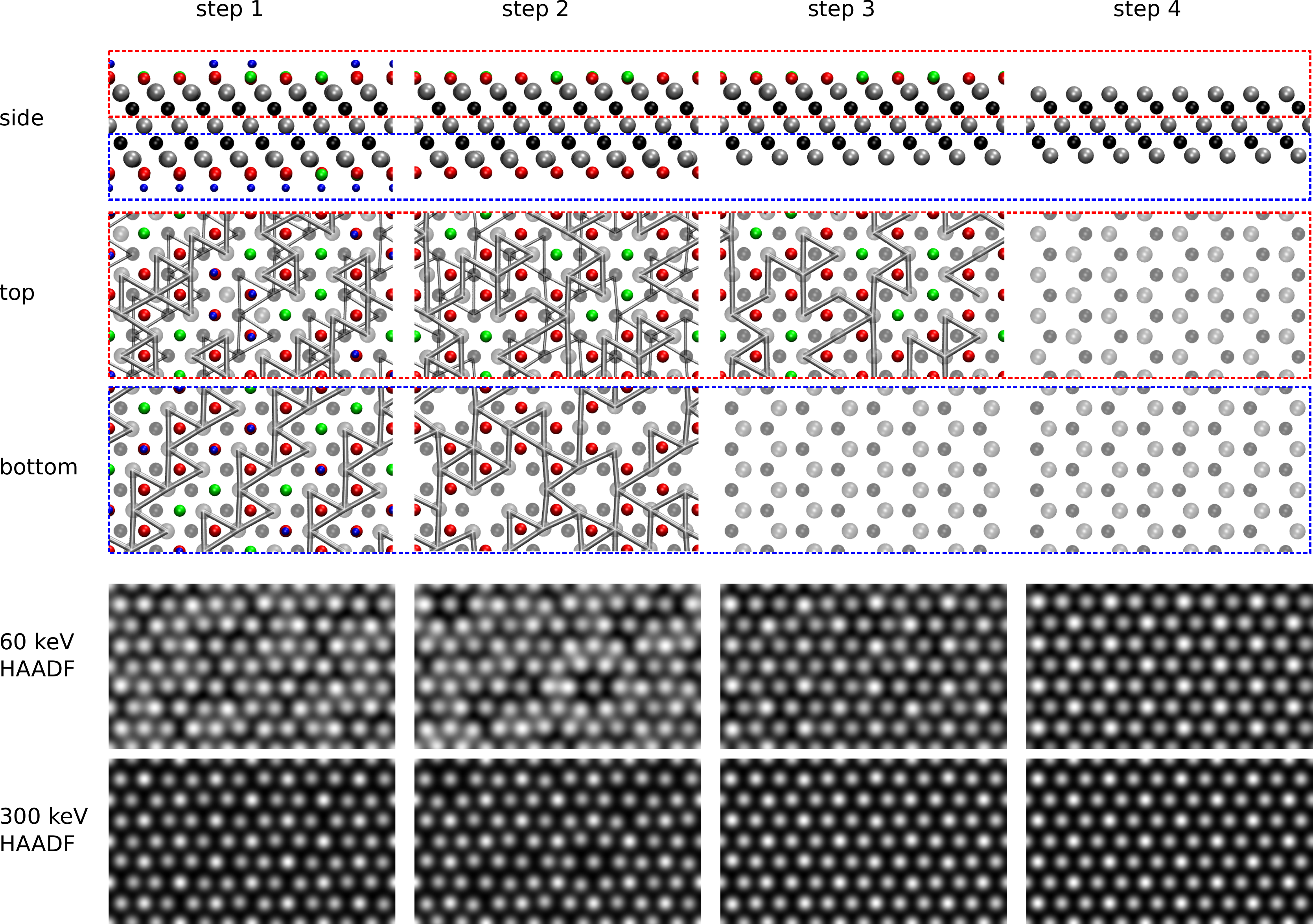}
\caption{\label{fig:temimage} 
Simulated TEM images of 
Ti$_3$C$_2$O$_{0.5}$F$_{0.25}$OH$_{0.25}$ at successive stages
of functional group sputtering.
Step 1 corresponds to initial fully covered surface,
steps 2--3 to partially covered surfaces, and
step 4 to bare surfaces. See text for full description.
(a) Side views of the structure. Atoms are colored as in Fig.\ \ref{fig:setup}.
(b) Top views of top part of the sheet [red box in (a)]. 
Only Ti-Ti bonds shorter than 3 {\AA} are shown.
Thin bonds are those from bottom part of the sheet.
(c) Bottom views of the bottom part of the sheet.
Only bonds for the bottom part are shown.
(d,e) Simulated HAADF STEM images for 60 keV (d) and 300 keV (e)
acceleration voltage.
} 
\end{figure*}

The results are collected in Fig.\ \ref{fig:temimage} where we
show the images for four cases:
(1) initial state with fully functionalized surface,
(2) surface after 1 s of 60 kV irradiation leading to bottom 
surface composition of O$_{0.75}$ and top surface composition of
O$_{0.75}$F$_{0.25}$,
(3) surface after 10s of 60 kV irradiation leading to bare bottom surface
and top surface composition of O$_{0.75}$F$_{0.75}$,
and (4) surface after 10s of 300 kV irradiation leading to two bare surfaces.


We first focus on the images simulated for the 300 kV acceleration voltage.
In step 1, the sites with two functionalized groups and one Ti atom per column show up brighter than the other two sites with only one Ti and one C atom per column. One cannot distinguish between O, OH, and F groups from the intensities.
Image from step 2 shows distinguishable intensity differences between functionalized and unfunctionalized sites on the bottom surface, but in step 3 the functionalized and unfunctionalized sites cannot be distinguished on the top surface.
Note, that step 4 with two bare surfaces still shows brighter spots, but not on the sites of surface functionalization as in steps 1--2. 
Instead, the bright spots occur at sites in which Ti atoms are located on the top surface and shows that the image contrast depends on the beam propagation direction and on the focus. The corresponding annular bright field images and a focus series are shown in the Supplementary information \cite{supp}.

Images simulated for the 60 kV acceleration voltage, in steps 1--2, show areas with darker and brighter contrast. 
These features do not arise from the brightness of the spots, but rather from the relative distance between the spots. I.e., the bright areas contain O, which locally contracts the lattice around it and leads to shorter Ti-Ti bonds, as illustrated in the top and bottom views of Fig.\ \ref{fig:temimage}.
In the 60 kV and 300 kV images the underlying structure is the same, but due to the larger extent of the spots in the 60 kV images, the relative distances are visually accentuated. 
On the other hand, this masks the brightness variations between the functionalized and unfunctionalized surfaces, that are perhaps easier to distinguish in the 300 kV images.
In steps 3 and 4 of 60 kV images, one again mostly sees the top-most Ti layer.

Given the experimental noise and any extraneous species adsorbed on the MXene surfaces, the identification of the surface groups is likely much more difficult than suggested by these simulated images.
Nevertheless, we think that the Ti-Ti bond contraction around O groups should be observable under the right imaging conditions and/or following suitable image processing, thus providing information about the composition and distribution of O on the MXene surfaces.

\subsection{Protecting Ti$_3$C$_2$ with monolayer graphene}

We also investigated the possibility of protecting the MXenes from beam damage by sandwiching it between graphene layers. 
Such strategy has been successfully employed to protect other 2D materials, such as transition metal dichalcogenides and black phosphorus 
\cite{AlgaraSiller13_APL,Nguyen17_ACSNano,Clark18_NL}.

\begin{figure}[h!]
\centering
\includegraphics[width=0.9\columnwidth]{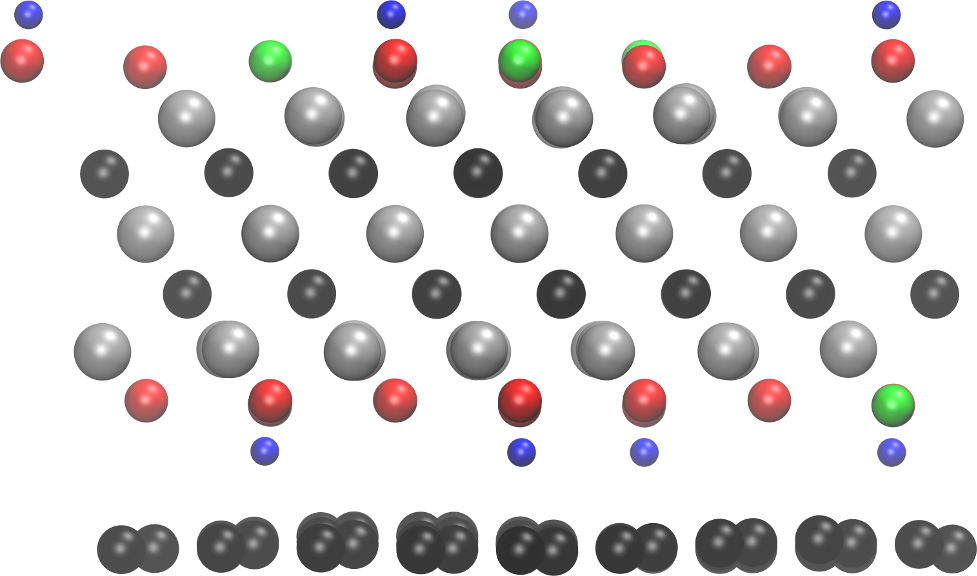}
\caption{\label{fig:hetero} 
The relaxed atomic structure for the
Ti$_3$C$_2$O$_{0.5}$F$_{0.25}$OH$_{0.25}$-graphene heterostructure.
The atoms are colored as in Fig.\ \ref{fig:setup}.} 
\end{figure}

The relaxed atomic structure is shown in Fig.\ \ref{fig:hetero}.
MXene layer remains flat, but there are pronounced corrugations in the graphene layer. These arise from the different interactions with the surface groups, i.e., shorter interlayer distance in the case of OH groups and larger in the case of O and F groups. 
Since these calculations become computationally much more demanding, we here carry out the simulations for only one electron energy, namely 100 keV,
and only for the bottom surface.
As can be seen in Table \ref{tab:en_mix}, at 100 keV all functional groups are expected to sputter easily from unprotected surface.

The graphene protects the bottom surface rather well from sputtering events.
At 100 keV all other surface groups remain stable, except H, which can be sputtered from the surface through the graphene lattice.
This is not surprising, given that the maximum transferred kinetic energy from 100 keV electron to H atoms is about 240 eV and the calculated energy barrier for atomic H penetrating graphene is only 2.6--4.6 eV \cite{hydrogen1,hydrogen2,hydrogen3}.
With such contrast of energies, it seems likely that H atoms could also be sputtered through graphene-protected top surface.
Thus, a heterostructure of MXenes without any OH groups and graphene
could be obtained by irradiating the sandwich structure under electron beam.

\subsection{Evaluated thresholds for other MXenes}

\begin{figure}[h!]
\centering
\includegraphics[width=\columnwidth]{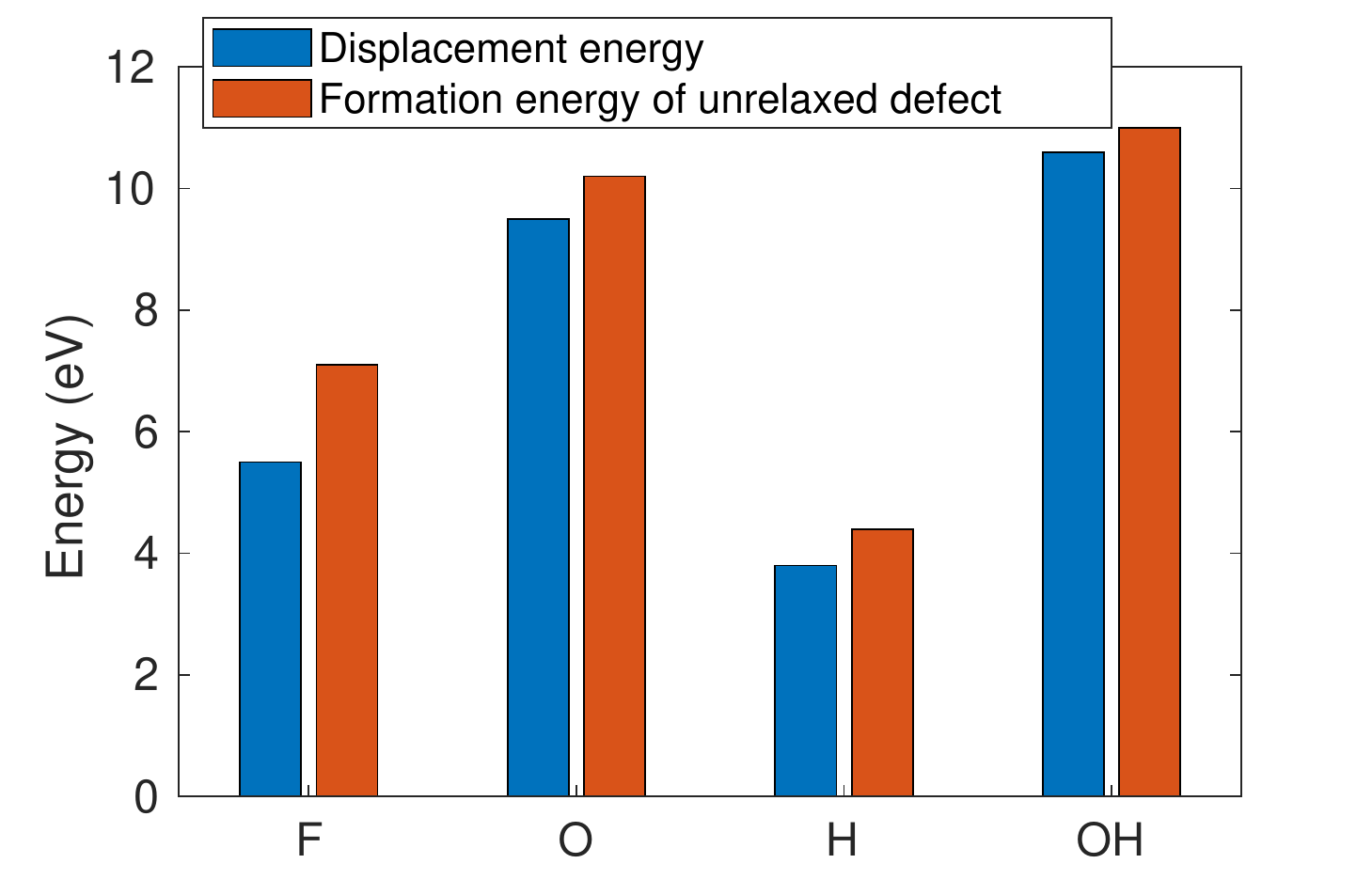}
\caption{\label{fig:thresholdbar} Displacement threshold energy of Ti$_3$C$_2$O$_{0.5}$F$_{0.25}$OH$_{0.25}$ surface vs.\ the formation energy of unrelaxed defects.}
\end{figure}

According to Ref.\ \cite{Komsa12_PRL}, the defect formation energies for unrelaxed defects are close to sputtering threshold kinetic energies from MD calculations in the case of "rigid" 2D materials. 
To verify that this holds true also for MXenes, we compare in Fig.\ \ref{fig:thresholdbar} the AIMD calculated displacement thresholds to the defect formation energies. The formation energy is consistently a bit higher (0.5--1 eV) than the displacement threshold, reflecting that small part of the energy is  deposited to the host. Note, that this approximation only works for the bottom atoms for which the sputtering trajectory is unobstructed. This allows us to qualitatively evaluate the threshold energies for many more MXene systems. 

The calculated defect formation energies for several distinct MXene systems such as Ti$_3$C$_2$, Ti$_2$N, Nb$_2$C, Mo$_2$TiC$_2$, and Ti$_3$CN  are  given in Table \ref{tab:form_en}. 
In this case, the metal atom vacancies are created in the bare surface, and O, F, and H vacancies in the pure O-, F-, and OH-covered surfaces.
Carrying out the whole procedure to determine the surface group composition and distribution for all different materials is beyond the scope of this paper.
The results are very similar for all Ti-containing MXenes. In the case of Nb$_2$C and Mo$_2$TiC$_2$, the metal atom sputtering threshold is somewhat higher and the O and F thresholds somewhat lower, but still differing by only about 1 eV.
From this, we conclude that bonding strength with surface groups is similar, and consequently the sputtering under electron microscopy should be similar for all MXenes, i.e., to the first approximation, our results for Ti$_3$C$_2$ should also be valid for other MXenes. It is also worth reminding that since Mo and Nb atoms are also clearly heavier than Ti, the transferred kinetic energy is lower and thereby the corresponding electron energy is much higher. Consequently Nb and Mo sputtering should be unlikely under typical acceleration voltages.

\begin{table}[h]
\centering
\caption{Formation energies for unrelaxed defects E$_{f}$ (in eV), and used to approximate T$_{k}$, for Ti$_3$C$_2$, Ti$_2$N, Nb$_2$C, Mo$_2$TiC$_2$, Ti$_3$CN. M refers to the outer metal atom, which is Mo in the case of Mo$_2$TiC$_2$.
} 
\label{tab:form_en} 
\begin{ruledtabular}
\begin{tabular}{lcccccc}

  & Ti$_3$C$_2$ & Ti$_2$C & Ti$_2$N & Nb$_2$C & Mo$_2$TiC$_2$ & Ti$_3$CN \\ \hline
M & 11.4      &11.4 & 10.8   & 14.0   & 15.3        & 11.0    \\
O & 10.2        & 10.2 & 10.1   & 9.8    & 8.7          & 10.0\\
F& 6.4&6.4&6.2&5.0&4.9&6.4\\
H&3.1&3.0&3.0&2.8&3.1&3.0\\
\end{tabular}
\end{ruledtabular}
\end{table}

\section{Conclusions}

We have carried out first principles calculations to study the stability of MXenes under electron beam.
In particular, the threshold energies for sputtering of surface group atoms via knock-on mechanism were evaluated via \emph{ab initio} molecular dynamics simulations.
We estimated the sputtering rates and also simulated the evolution of the surface group composition over time.
It was found that the bottom surface can be selectively cleared of the surface groups when using low acceleration voltages (60 kV). At high voltages (300 kV), also the surface groups of the top surface start to sputter.
After clearing the surface groups, Ti atoms on the bare surface are relatively stable under the microscope.
Since the bare surface is highly reactive, it can adsorb  almost anything that is introduced afterwards, e.g., CO$_2$ \cite{Persson19_AM}, thus paving a way for engineering of the MXene surface. On the other hand, this also suggests that any residual gases in the chamber are likely to stick to the surface and hinder the imaging.
Our simulated TEM images revealed that it might be possible to identify the degree of cleaning, i.e., whether there are surface groups or not, but identification of the type of surface groups is not possible based only on the intensity. However, the O groups lead to contraction of the lattice and shortening of the Ti-Ti bonds that is likely to be observable.
In addition, we propose that graphene encapsulation could provide a viable pathway for protecting the MXene layers from electron beam damage during imaging, except for H which can be sputtered out through the graphene sheets.
Finally, we evaluated the sputtering thresholds for five other MXenes using formation energies of unrelaxed defects and found them to be close to those found for Ti$_3$C$_2$ and we thus believe that our findings for Ti$_3$C$_2$ are also largely valid for MXenes more generally.

\section*{Acknowledgments}

We acknowledge funding from the Academy of Finland under Project No. 311058. 
We thank CSC Finland for generous grants of CPU time. We also want to thank Per Persson and Ingemar Persson for discussions and comments on the manuscript.

\bibliography{defects}

\end{document}


\author{Rina Ibragimova$^1$}
\author{Zhong-Peng Lv$^1$}
\author{Hannu-Pekka Komsa$^{1,2}$}
\affiliation{
$^1$Department of Applied Physics, Aalto University, 
Finland
}
\affiliation{
$^2$Microelectronics Research Unit, University of Oulu, 
Finland}

\title{Supplementary information for "First principles study of stability of MXenes under electron beam"}

\maketitle

\section{Rate equations}

Let us denote the number of functional groups (per surface site) as 
$N_O$, $N_{OH}$, and $N_F$, and $N_{vac}$ for empty site. Assuming these are the only possible ones, then $N_O + N_{OH} + N_F + N_{vac} = 1$.
The rate equation for F is simply given by the corresponding sputtering rate $r_F$, but the numbers for O and OH are coupled since sputtering of H from OH give O group:
\begin{align}
\frac{dN_F}{dt} &= -r_F N_F \\
\frac{dN_{OH}}{dt} &= -r_{OH} N_{OH} - r_H N_{OH} \\
\frac{dN_O}{dt} &= -r_O N_O + r_H N_{OH}
\end{align}
Using a shorthand notation $r' = r_H + r_{OH}$
the analytic solutions for these are:
\begin{align}
N_F &= N_F^0 e^{-r_Ft} \\
N_{OH} &= N_{OH}^0 e^{-r't} \\
N_O &= \frac{r_H N_{OH}^0}{r_O-r'}e^{-r' t} +
    (N_O^0 - \frac{r_H N_{OH}^0}{r_O-r'})e^{-r_O t}
\end{align}


\section{Simulated annular bright field images}

\begin{figure*}[h!]
\centering
\includegraphics[width=\textwidth]{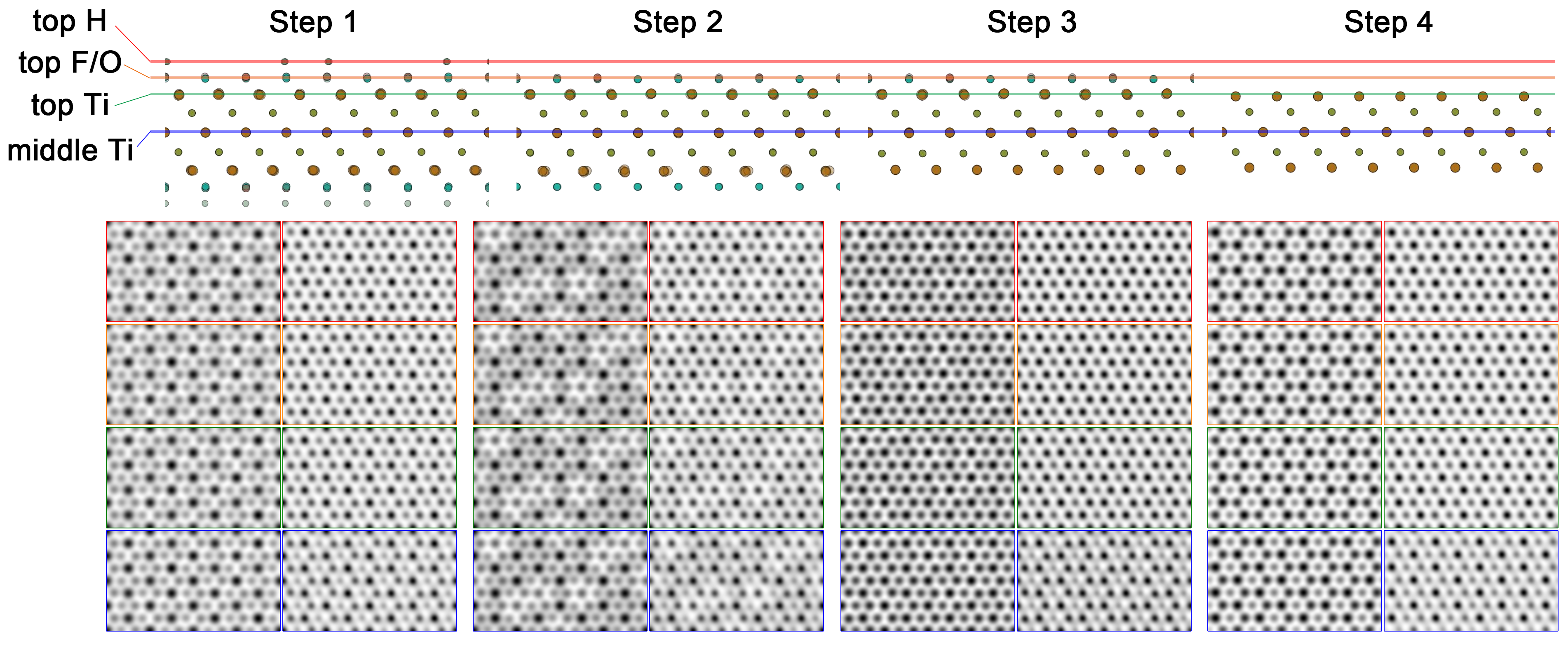}
\caption{\label{fig:defocus} 
Annular bright field images for the four structures
considered in the main paper at 60 keV (left) and 300 keV (right) acceleration voltages and four different focus points (indicated by the box color).
}
\end{figure*}

Annular bright field (ABF) images for the same structures
and energies as in Fig. 4 of the main paper are
shown in Fig.\ \ref{fig:defocus}.
In addition, it shows how the images depend on the focus point.

\bibliography{defects}